\documentclass[twocolumn, prl, showpacs,superscriptaddress]{revtex4}
\usepackage{amssymb}
\usepackage{graphicx}
\usepackage{dcolumn}
\usepackage{bm}
\usepackage{amsmath}

\begin{document}

\title{Steady-state localized-delocalized phase transition of an
incoherent-pumped dissipative Bose-Hubbard model}
\author{Yuanwei Zhang}
\affiliation{Beijing National Laboratory for Condensed Matter Physics, Institute of
Physics, Chinese Academy of Sciences, Beijing 100190, China}
\affiliation{School of Physical Sciences, University of Chinese Academy of Sciences,
Beijing 100190, China}
\author{Jingtao Fan}
\affiliation{State Key Laboratory of Quantum Optics and Quantum Optics Devices, Institute
of Laser Spectroscopy, Shanxi University, Taiyuan, Shanxi 030006, China}
\affiliation{Collaborative Innovation Center of Extreme Optics, Shanxi University,
Taiyuan, Shanxi 030006, China}
\author{Gang Chen}
\thanks{chengang971@163.com}
\affiliation{State Key Laboratory of Quantum Optics and Quantum Optics Devices, Institute
of Laser Spectroscopy, Shanxi University, Taiyuan, Shanxi 030006, China}
\affiliation{Collaborative Innovation Center of Extreme Optics, Shanxi University,
Taiyuan, Shanxi 030006, China}
\author{Wu-Ming Liu}
\thanks{wmliu@iphy.ac.cn}
\affiliation{Beijing National Laboratory for Condensed Matter Physics, Institute of
Physics, Chinese Academy of Sciences, Beijing 100190, China}
\affiliation{School of Physical Sciences, University of Chinese Academy of Sciences,
Beijing 100190, China}

\begin{abstract}
We investigate steady-state properties of a two-dimensional
incoherent-pumped dissipative Bose-Hubbard model, which describes a photon
square lattice. This incoherent pumping exhibits an important
environment-induced higher-order fluctuation effect, which induces a strong
competition between the driven-dissipative channel, the photon-photon
interaction, and the photon hopping in multi-photon processes. This new
competition gives rise to a spontaneous breaking of the $U(1)$ symmetry of
system. As a result, we predict a many-body steady-state
localized-delocalized phase transition and an anti-blockade effect, in which
the increasing of the repulsive photon-photon interaction promotes the
emergence of phase transition. These unconventional many-body steady-state
phenomena can be understood by analyzing the single-cavity properties. Our
results pave a new way to control many-body dynamics of driven-dissipative
systems.
\end{abstract}

\maketitle

\textit{Introduction.}--Understanding and controlling quantum many-particle
systems is an fundamental task but a grand challenge in modern physics. A
crucial problem is that almost all many-particle systems are coupled to the
environment and thus are subjected to unavoidable dissipations, which are
usually compensated by external incident laser fields. Recently, both
theoretical \cite{SD08,FV09,HW10,SD10,DDBR13,AWC13,BB15,MA16,MZ16,FR16} and
experimental \cite{HK11,JTB11,PS13} works have demonstrated that the
dissipations can create correlations between particles and represents the
dominate resource of many-body dynamics. Following this pioneer discovery,
driving-dissipative many-body systems have attracted great attention, and
especially, new many-body correlation characters far from thermal
equilibrium have been revealed \cite%
{SD10-2,AT11,SD11,CE12,FN12,ALB13,LMS13,JJ14,MM14,LB14,HW15,JC15,SF15,EM15,LMS15,DN15,GD15,NL15,CKC15,JCB15,AHW16,AC16,JJ16,JM16,RL16,BE16,JJM16,MFM16,MS16,EL16}%
.

In addition to the dissipation processes, the coupling with the environment
also induces actual random fluctuating driven processes. In thermal
equilibrium case, they are not independent but completely determine each
other through the fluctuation--dissipation theorem. This mechanism is the
origin of random phenomena, which are the key to understand the complexity
of real word. In classical case, the most famous example is the Brownian
motion of particles \cite{Feynman}. And in quantum scale, these effects have
been proved to play an important role, for example, in determining the
cooling limit of optomechanical systems \cite{MA14}. In general, the random
fluctuating force induces the thermal noise, which is harmful for studying
quantum phase transitions and should be inhibited by reducing the
temperature of the environment. In this Letter, we reveal that the
fluctuating driven processes can generate exotic steady-state phases and
phase transitions in non-equilibrium systems.
\begin{figure}[tb]
\includegraphics[width=8.5cm]{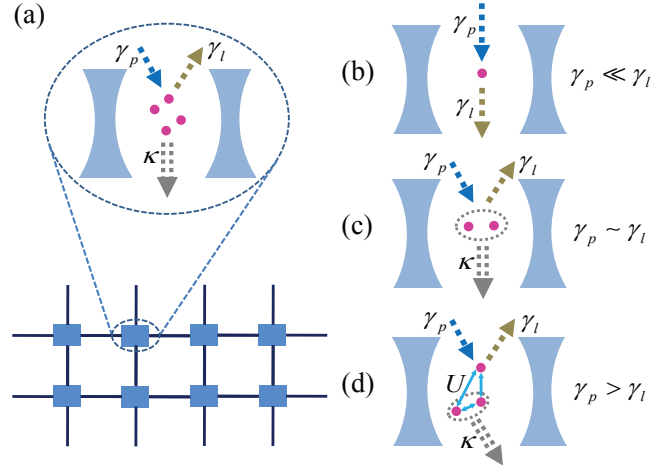}\newline
\caption{(a) Sketch of a photon square lattice made of nonlinear cavities,
with the incoherent-pumped process $\protect\gamma _{p}$, the single-photon
loss process $\protect\gamma _{l}$, and the two-photon loss process $\protect%
\kappa $. (b)-(d) Schematics of the single-cavity driven-dissipative
processes for different parameters, such that (b) $\protect\gamma _{p}\ll
\protect\gamma _{l}$, (c) $\protect\gamma _{p}\sim \protect\gamma _{l}$, and
(d) $\protect\gamma _{p}>\protect\gamma _{l}$.}
\label{fig1}
\end{figure}

To show our arguments, we consider a two-dimensional (2D) incoherent-pumped
dissipative Bose-Hubbard (BH) model \cite%
{Carusotto,JK06,DB08,SIT09,Keeling07}, which describes a photon square
lattice shown in Fig.~\ref{fig1}(a) \cite{Hartmann}. The incoherent-pumped
process, realized in experiments \cite%
{Carusotto,JK06,DB08,SIT09,Keeling07,Houck}, is a kind of random driven
process. Here, this incoherent pumping induces a strong competition between
the driven-dissipative channel, the photon-photon interaction, and the
photon hopping in multi-photon processes. This new competition gives rise to
a spontaneous breaking of the $U(1)$ symmetry of system. As a result, we
predict a many-body steady-state localized-delocalized phase transition and
an anti-blockade effect \cite{blockade}, in which the increasing of the
repulsive photon-photon interaction promotes the emergence of phase
transition. We emphasize that the steady-state phenomena predicted can be
understood by analyzing the single-cavity properties, and have no
correspondence in equilibrium case. Moreover, they are fully governed by the
multi-photon processes, arising from the environment-induced higher-order
fluctuations. However, in the mean-field level these multi-photon processes
are usually neglected and the relevant physics cannot be captured. To
overcome this shortcoming, we introduce a non-equilibrium Green's function
approach.

\textit{Incoherent-pumped dissipative Bose-Hubbard model.}--The
incoherent-pumped dissipative dynamics of the 2D BH model is governed by the
following Lindblad master equation of the many-body density matrix $\rho
\left( t\right) $ ($\hbar =1$ hereafter) \cite{Carusotto}:%
\begin{eqnarray}
\partial _{t}\rho &=&-i\left[ H,\rho \right] +\gamma _{p}\sum_{i}\left(
a_{i}^{\dag }\rho a_{i}-\frac{1}{2}\left[ a_{i}a_{i}^{\dag },\rho \right]
_{+}\right)  \notag \\
&&+\gamma _{l}\sum_{i}\left( a_{i}\rho a_{i}^{\dag }-\frac{1}{2}\left[
a_{i}^{\dag }a_{i},\rho \right] _{+}\right)  \notag \\
&&+\kappa \sum_{i}\left( a_{i}a_{i}\rho a_{i}^{\dag }a_{i}^{\dag }-\frac{1}{2%
}\left[ a_{i}^{\dag }a_{i}^{\dag }a_{i}a_{i},\rho \right] _{+}\right) ,
\label{L1}
\end{eqnarray}%
with
\begin{equation}
H=-\frac{J}{z}\sum_{\langle i,j\rangle }a_{i}^{\dag }a_{j}+\omega
_{c}\sum_{i}a_{i}^{\dag }a_{i}+\frac{U}{2}\sum_{i}a_{i}^{\dag }a_{i}^{\dag
}a_{i}a_{i}.  \label{BH}
\end{equation}%
In Eq.~(\ref{L1}), $\left[ \cdots \right] _{+}$ is the anticommutator, $%
a_{i}^{\dag }$ creates a photon on site $i$, $\gamma _{p}$ describes the
incoherent-pumped process, and $\gamma _{l}$ and $\kappa $ govern
respectively the single- and two-photon loss processes \cite{INT}. In the BH
Hamiltonian (\ref{BH}), $\langle i,j\rangle $ indicates that the photon can
hop between adjacent cavities, $J>0$ is the hopping strength, $z=4$ is the
coordination number, $\omega _{c}$ is the photon frequency, and $U>0$
represents the repulsive photon-photon interaction \cite{INT}.

Since the injected photons from the incoherent pumping obey random
distribution, Eq.~(\ref{L1}) still holds a global $U(1)$\ symmetry. More
interestingly, the environment-induced higher-order fluctuations induce
multi-photon processes, in which the driven-dissipative channel has a strong
competition with the photon-photon interaction and the photon hopping. This
new competition gives rise to a spontaneous breaking of the $U(1)$\
symmetry, and thus a many-body steady-state phase transition, from a
localized state (LS) to a delocalized (superfluid) state (DS), is expected
to emerge \cite{U1}.

To better understand relevant physics, we begin to qualitatively analyze a
single-cavity problem. When $\gamma _{p}\ll \gamma _{l}$, the single-photon
loss process makes the injected photons decay to the environment very fast;
see Fig.~\ref{fig1}(b). When $\gamma _{p}\sim \gamma _{l}$, the probability
of photons staying in the cavity becomes larger and two photons maybe appear
at the same time. Hence, the two-photon loss channel opens; see Fig.~\ref%
{fig1}(c). When $\gamma _{p}>\gamma _{l}$, there exists an effective gain of
photons through the single-particle process. In this case, multiple photons
maybe appear in the cavity and the probability of the two-photon loss event
is increased. As a result, the effective gain process is balanced by the
two-photon loss process; see Fig.~\ref{fig1}(d). Interestingly, in this
region the photon-photon interaction plays a crucial role in determining the
systematic properties. Especially, when it is strong, the photons oscillate
faster and the relative possibility of the dissipative events is decreased.
It implies that the photon-photon interaction not only governs the
excitation energy of the multi-photon state, but also increases the
linewidth of excitation \cite{Note}. If the linewidth is divergent, the
steady state, with random distribution of multiple photons discussed above,
becomes unstable, and thus a new steady state emerges, since the
photon-photon interaction makes the photons tend to oscillate with a uniform
phase.

For the photon square lattice, the many-body steady state is the result of
the detailed balance between the photon input and output processes of each
cavity. It refers to not only the driven-dissipative processes but also the
photon hopping between adjacent cavity. When the hopping strength becomes
strong enough, the photons can hop in all lattices without decay into the
environment. Therefore, a steady-state phase transition, from the LS to the
DS, occurs. However, in the mean-field level these multi-photon processes discussed above
are usually neglected and the relevant many-body physics cannot be captured
\cite{SM1}. To overcome this shortcoming, here we introduce a
non-equilibrium Green's function approach.

\textit{Noise state of the single cavity}.--Similar to the previous
qualitative analysis, we also begin to quantitatively consider a
single-cavity case, in which $a_{i}$ is replaced by $a$. We mainly capture
its single-particle excitation spectra by calculating the retard Green's
function $G_{0}^{R}\left( t\right) =-i\theta \left( t\right) \left\langle %
\left[ a\left( t\right) ,a^{\dag }\left( 0\right) \right] \right\rangle $
\cite{Note1}, where $\theta \left( t\right) $ is the Heaviside step
function. A simple way to obtain $G_{0}^{R}\left( t\right) $\ is taken into
account its dynamics, $i\dot{G}_{0}^{R}\left( t\right) =\delta \left(
t\right) -\theta \left( t\right) \left\langle \left[ \left( i\omega
_{c}-\chi \right) a\left( t\right) ,a^{\dag }\left( 0\right) \right]
\right\rangle -\theta \left( t\right) \left\langle \left[ \left( iU+\kappa
\right) a^{\dag }a^{2}\left( t\right) ,a^{\dag }\left( 0\right) \right]
\right\rangle $ \cite{SM2}, where $\delta \left( t\right) $ is the delta
function and $\chi =\left( \gamma _{p}-\gamma _{l}\right) /2$ describes the
effective gain of photons through the single-particle process.

The term $-i\theta
\left( t\right) \left\langle \left[ a^{\dag }a^{2}\left( t\right) ,a^{\dag
}\left( 0\right) \right] \right\rangle $ is called the second-order time-ordered correlation function, and can be obtained by the
higher-order terms through general recursive relations \cite{SM2}. It reflects
the important environment-induced quantum fluctuation effect. Due to the existence of this term,
the above dynamical equation is not a closed equation. In the mean-field level, this equation can be linearized and becomes a closed equation. Unfortunately, under this approximation the relevant
physics will be lost \cite{SM2}. In the following discussions, we carefully
consider the higher-order time-ordered correlation functions.

\begin{figure}[t]
\includegraphics[width=8.4cm]{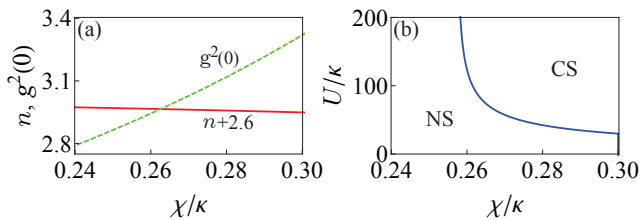}\newline
\caption{(a) $n$ and g$^{2}(0)$ in the noise state versus $\protect\chi $.
(b) Steady-state phase diagram of the single cavity versus $U$ and $\protect%
\chi $. Here, NS and CS denote the noise and coherent classical states, respectively. In
these figures, $\protect\gamma _{p}/\protect\kappa =0.60$ and the maximum of
$\protect\chi /\protect\kappa $ is thus $0.30$, according to the definition.}
\label{fig2}
\end{figure}

In current experiments \cite{EIT}, the maximal photon number in the cavity
is small. We assume that there are at most four incident photons, and then
obtain the retard Green's function after neglecting the time-ordered
correlation functions higher than the fourth order. We find that when $U$ is
weak, there exists a novel noise state, characterized by $\left\langle
a\right\rangle =0$\ and $\langle a^{\dag }a\rangle =n\neq 0$, which clearly
demonstrates the random distribution of the photon field. It is quite
different from the mean-field prediction \cite{SM1}, with a vacuum state ($%
\left\langle a\right\rangle =0$ and $n=0$). In Fig.~\ref{fig2}(a), we plot
the mean-photon number $n$\ and the second-order correlation function g$%
^{2}(0)=\langle a^{\dag 2}a^{2}\rangle /\langle a^{\dag }a\rangle ^{2}$ to
show the features of the photon field distribution in the noise state. When
increasing $U$, a steady-state phase transition, from a noise state to a
coherent classical state ($\left\langle a\right\rangle \neq 0$\ and $n\neq 0$%
), can be predicted \cite{SM2}. In Fig.~\ref{fig2}(b), we plot the
corresponding phase diagram versus $U$ and $\chi $ \cite{Note2}.
\begin{figure}[b]
\includegraphics[width=8.0cm]{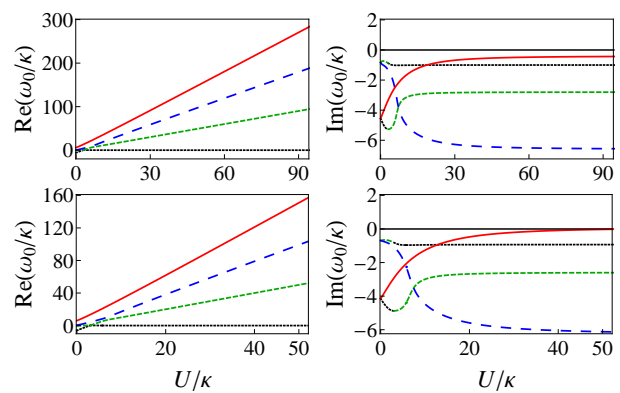}\newline
\caption{The single-particle excitation spectra of the single cavity versus $%
U$ for $\protect\chi /\protect\kappa =0.20$ (upper panel) and $\protect\chi /%
\protect\kappa =0.27$ (lower panel), with $\protect\gamma _{p}/\protect%
\kappa =0.60$. Since there are at most four incident photons in the cavity,
the excitation spectra have four branches, which are labeled by different
linetypes and colors.}
\label{fig3}
\end{figure}

As shown in Fig.~\ref{fig2}(a), when increasing $\chi $, $n$ is not
increased, which means that the effective input photons are dissipated to
the environment through the two-photon loss process. On the contrary, g$%
^{2}(0)$ is increased, which implies that the environment-induced quantum
fluctuations are enhanced, i.e., the probability of the multi-photon events
becomes larger. As we discussed in qualitative analysis, the repulsive
interaction between multiple photons will make the noise state become
unstable. Therefore, the phase transition only occurs for a strong $\chi $,
as shown in Fig.~\ref{fig2}(b), and moreover, the critical photon-photon
interaction strength $U_{c}$ is decreased when increasing $\chi $. In
contrast, when $\chi <0.26$, $U_{c}$ is rapidly increased to infinity and
thus the phase transition could not happen. To gain deeper insight of phase
transition, in Fig.~\ref{fig3} we plot the single-particle excitation
spectra $\omega _{0}$, determined by the poles of the retarded Green's
function in the frequency space. The real parts of $\omega _{0}$,
abbreviated as Re$\left( \omega _{0}\right) $, reflect the excitation
energies, and the imaginary parts of $\omega _{0}$, abbreviated as Im$%
(\omega _{0})$, govern the linewidths of the excitation spectra. When $U$ is
large enough, the excitation energies are given approximately by $0$, $U$, $%
2U$, and $3U$, which correspond to the change of the total photon-photon
interaction energy when adding or removing one photon. We also note that
when increasing $U$, one branch of Im$(\omega _{0})$ (the red line in Fig.~%
\ref{fig3}), which corresponds to excitation of the multi-photon state, is
increased. If $\chi $ is small ($\chi =0.20$), it becomes a negative
constant. On the contrary, when $\chi $ is large ($\chi =0.27$), it will
reach $0$ for a strong $U$. This means the lifetime of excitation is
divergent, and thus the noise state becomes instable.

It should be emphasized that the noise state and the properties of the
excitation spectra of the single cavity are crucial for exploring and
understanding the many-body steady-state phase transition for the 2D
incoherent-pumped dissipative BH model, governed by the Lindblad master
equation (\ref{L1}).

\begin{figure}[b]
\includegraphics[width=8.4cm]{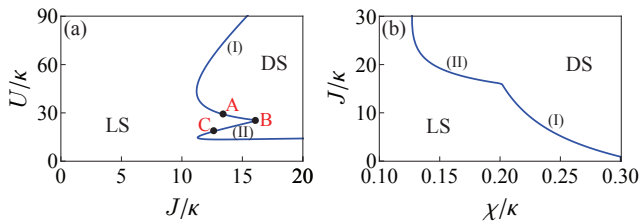}\newline
\caption{(a) Many-body steady-state phase diagram versus $J$ and $U$ for $%
\protect\chi /\protect\kappa =0.2$ and $\protect\gamma _{p}/\protect\kappa %
=0.6$. A, B, and C are the chosen points, whose different excitation spectra
are presented in Fig.~5. (b) Phase diagram versus $J$ and $\protect\chi $
for $U/\protect\kappa =25$ and $\protect\gamma _{p}/\protect\kappa =0.6$. In
these figures, (I) and (II) label the different kinds of the phase
transition, from the localized state (LS) and delocalized state (DS).}
\label{fig4}
\end{figure}

\textit{Many-body steady-state phase transition.}--To explore the many-body
steady-state properties of Eq.~(\ref{L1}), we first assume that every cavity
is initially prepared in its noise state, which means that the photons are
localized at each site. Then, we introduce the Keldysh functional-integral
formalism to calculate the full retard Green's function $\tilde{G}^{R}\left(
t\right) =-i\left\langle a_{i,cl}(t)a_{i,q}^{\ast }(0)\right\rangle $ \cite%
{AB10}, which is dressed by the hopping terms. In details, we use a
non-equilibrium linked-cluster expansion approach, which gives a description
of equilibrium or non-equilibrium strong correlation systems within the same
formalism \cite{AB10,Pelster,Kennett}. Following this method, all the
single-site terms are regarded as the unperturbed parts and the hopping
terms are treated as perturbations \cite{SM3}.

\begin{figure}[t]
\includegraphics[width=8.0cm]{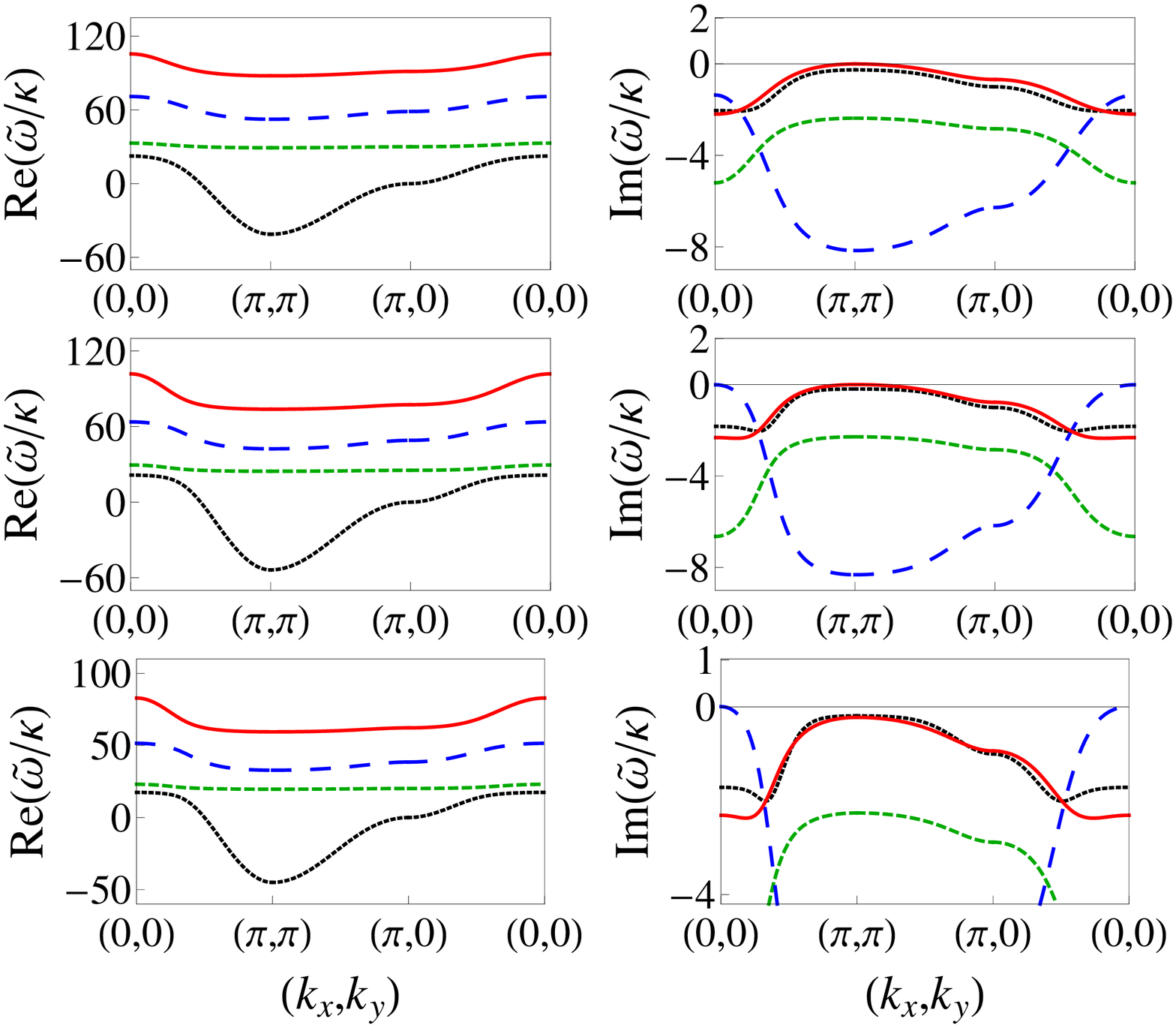}\newline
\caption{Energy-momentum dispersions of elementary excitations for points A
(upper panel), B (middle panel), and C (lower panel), indicated in
Fig.~4(a). $(0,0)$, $(\protect\pi ,\protect\pi )$, and $(\protect\pi ,0)$
are the special points in the Brillouin zone of the square lattice. Here,
the elementary excitations have four branches, which are labeled by
different linetypes and colors.}
\label{fig5}
\end{figure}

We sum an infinite set of diagrams by calculating the irreducible part of
the retard Green's function $K^{R}\left( t\right) $, which is connected, in
the frequency space, to the full Green's function via the following
equation: $\tilde{G}^{R}\left( \mathbf{k},\omega \right) =K^{R}\left(
\mathbf{k},\omega \right) /\left[ 1-J\left( \mathbf{k}\right) K^{R}\left(
\mathbf{k},\omega \right) \right] $, with the 2D lattice dispersion $J(%
\mathbf{k})=2J\cos (\mathbf{k}\cdot \mathbf{r})$, where $\mathbf{k}$\ is the
wave vector and $\mathbf{r}$\ is the lattice vector, with $\left\vert
\mathbf{r}\right\vert =1$. We mainly consider the contribution of the chain
diagram to $K^{R}\left( t\right) $. By setting $K^{R}\left( \omega \right)
=G_{0}^{R}\left( \omega \right) $, where $G_{0}^{R}\left( \omega \right) $
is the single-cavity retard Green's function in the frequency-momentum
space, we can obtain a non-equilibrium Dyson equation about the inverse of
the full retard Green's function, i.e.,
\begin{equation}
\left[ \tilde{G}^{R}\left( \mathbf{k},\omega \right) \right] ^{-1}=\left[
G_{0}^{R}\left( \omega \right) \right] ^{-1}-\Sigma ^{R}\left( \mathbf{k}%
,\omega \right) ,  \label{DE}
\end{equation}%
where $\Sigma ^{R}\left( \mathbf{k},\omega \right) =J(\mathbf{k})$\ is the
self-energy \cite{SM3}.

Equation~(\ref{DE}) is the main result of this Letter. Although this
equation is similar to that in equilibrium case, the undertaken physics is
quite different, because it contains all the driven-dissipative terms. For
equilibrium case, a many-body phase transition, whose phase boundary is
characterized by the free energy, emerges. Whereas for non-equilibrium case
considered here, we donot have a sensible notion of a free energy and a
many-body steady-state phase transition is expected to occur. Moreover, the
corresponding phase boundary is completely determined by the characteristic
frequencies $\tilde{\omega}$\ of the single-particle excitation spectra,
which are the poles of $\tilde{G}^{R}\left( \mathbf{k},\omega \right) $,
i.e., $[\tilde{G}^{R}\left( \mathbf{k},\tilde{\omega}\right) ]^{-1}=0$. When
the imaginary parts of $\tilde{\omega}$ are negative, a localized steady
state is stable, otherwise this state is instable and a delocalized steady
state emerges.

In Fig.~\ref{fig4}(a), we plot the many-body steady-state phase diagram
versus $U$\ and $J$, with the same driven-dissipative parameters as the
upper panel of Fig.~\ref{fig3}. For a weak $U$, the driven-dissipative
induced dephasing is dominant and no steady-state phase transition can be
found, as expected. When $U$ is strong, it suppresses the dephasing effect
and all linewidths of the single-cavity excitations become constant; see
Fig.~\ref{fig3}. Thus, the many-body steady-state LS-DS phase transition
occurs, and moreover, is dominated by the competition between $U$\ and $J$.
These predicted results are sharply contrast to those derived from the
mean-field level \cite{Note3}, in which the important environment-induced quantum fluctuations have been
neglected.

More interestingly, for an intermediate $U$, the single-cavity excitations
are very complex, and the many-body steady-state phase diagram exhibits
unconventional behaviors. For example, the phase boundary is consisted of
two smooth curves, which are labeled respectively as $\left( \text{I}\right)
$ and $\left( \text{II}\right) $ and connected together at a tip (point B).
We emphasize that these two curves reflect different features of phase
transition. In curve $\left( \text{I}\right) $, the instabilities arise at $%
k=(\pi ,\pi )$, corresponding to a multi-photon model (red line), while in
curve $\left( \text{II}\right) $, the instabilities arise at $k=(0,0)$,
corresponding to a photon-pair model (blue line). At the tip, the
instabilities arise from $k=(\pi ,\pi )$ and $k=(0,0)$ at the same time.
These properties can be confirmed in Fig.~\ref{fig5}, in which we plot the
energy-momentum dispersions of the elementary excitations for points A, B,
and C, indicated in Fig.~\ref{fig4}(a). These properties can be explained as
follows. When increasing the relatively small $U$, the linewidth of
excitation, which corresponds to the photon-pair model, is dramatically
decreased (see the blue line in Fig.~\ref{fig3}). This means that the
photon-photon interaction makes two photons tend to be oscillating with same
phases, which enhances the two-photon loss process. In this case, when
increasing $J$, the photons will hop into adjacent cavities if they have
same phases. As a result, this phase consistency promotes the emergence of
the many-body steady-state phase transition, with $k=(0,0)$. When $U$ is
large, the maximal Im$(\omega _{0})$ corresponds to the excitation of the
multi-photon state (see the red line in Fig.~\ref{fig3}). In this case, the
photons are more likely tunneling into adjacent cavities if they have
opposite phases to overcome the repulsive interaction. Thus, the
instabilities arise from $k=(\pi ,\pi )$. A similar phenomenon can be found
in Fig.~\ref{fig4}(b), in which when increasing $\chi $, the multi-photon
processes play a dominate role and the most instable mode changes from $%
k=(0,0)$\ to $k=(\pi ,\pi )$.

In addition, when increasing $U$ for a fixed $J/\kappa (=15)$, we observe
two LS-DS-LS phase transitions; see also Fig.~\ref{fig4}(a). The first one
shows that the DS only occurs when $U$\ is large enough as we discussed
above. When increasing $U$, the system goes back into the LS. This property
is attributed to the competition between $U$ and $J$ and reflects a photon
blockade effect. When further increasing $U$, the second LS-DS-LS phase
transition occurs. This means that there exists a anti-blockade effect, in
which the increasing of $U$\ promotes the emergence of phase transition. The
main reason is that when increasing $U$, the linewidth of the excitation of
the multi-photon state becomes longer (see the red line in Fig.~\ref{fig3})
and the photons are more easily to enter the adjacent cavities.

\textit{Conclusions.}--In summary, we have explored the non-equilibrium
physics of a 2D incoherent-pumped dissipative BH model, by introducing a
non-equilibrium Green's function approach. We have predicted a many-body
steady-state localized-delocalized phase transition and revealed an
interesting anti-blockade effect. We have shown that all unconventional
many-body steady-state features arise from the environment-induced
higher-order fluctuations and can be explained by analyzing the
single-cavity properties. Our results pave a new way to control many-body
dynamics of driven-dissipative systems.

We thank Prof.~Rosario Fazio, Prof.~Hendrik Weimer, and Dr.~Yu Chen for
numerous insightful discussions. This work was supported in part by the
NKRDP under Grants No.~2012CB821305 and No.~2016YFA0301500; the NSFC under
Grants No.~61227902, No.~61275211, No.~61378017, No.~11422433, No.~11434015,
and No.~11674200; SKLQOQOD under Grants No.~KF201403; SPRPCAS under Grants
No.~XDB01020300 and No.~XDB21030300; the FANEDD under Grant No.~201316;
OYTPSP; and SSCC.


\begin{thebibliography}{99}
\bibitem{blockade} In general, the strong interaction between the photons
causes a photon blockade of cavity transmission, i.e., the presence of
photons in the cavity can block the resonant injection of another photon
\cite{Carusotto}.
\bibitem{INT} A possible way to realize the photon-photon interaction is
used the electromagnetically-induced transparency configuration \cite%
{Carusotto}. In this case, the two-photon loss process can also appear
naturally and be tuned independently.
\bibitem{Note} In equilibrium case, the photon-photon interaction determines
the photon distribution by minimizing the free energy. However, in
nonequilibrium case we donot have a sensible notion of free energy, and the
energy spectrum is defined in complex plane. Its real and imaginary parts
reflect the excitation energy and linewidth, respectively.
\bibitem{EIT} In current experiments of Rydberg atoms about the multi-photon
electromagnetically-induced transparency \cite{Carusotto}, the maximal
photon number in the cavity is small and tunable.
\end{thebibliography}
\end{document}


\title{Supplementary Material ``Steady-state localized-delocalized phase
transition of an incoherent-pumped dissipative Bose-Hubbard model"}
\author{Yuanwei Zhang}
\affiliation{Beijing National Laboratory for Condensed Matter Physics, Institute of
Physics, Chinese Academy of Sciences, Beijing 100190, China}
\affiliation{School of Physical Sciences, University of Chinese Academy of Sciences,
Beijing 100190, China}
\author{Jingtao Fan}
\affiliation{State Key Laboratory of Quantum Optics and Quantum Optics Devices, Institute
of Laser Spectroscopy, Shanxi University, Taiyuan, Shanxi 030006, China}
\affiliation{Collaborative Innovation Center of Extreme Optics, Shanxi University,
Taiyuan, Shanxi 030006, China}
\author{Gang Chen}
\thanks{chengang971@163.com}
\affiliation{State Key Laboratory of Quantum Optics and Quantum Optics Devices, Institute
of Laser Spectroscopy, Shanxi University, Taiyuan, Shanxi 030006, China}
\affiliation{Collaborative Innovation Center of Extreme Optics, Shanxi University,
Taiyuan, Shanxi 030006, China}
\author{Wu-Ming Liu}
\thanks{wmliu@iphy.ac.cn}
\affiliation{Beijing National Laboratory for Condensed Matter Physics, Institute of
Physics, Chinese Academy of Sciences, Beijing 100190, China}
\affiliation{School of Physical Sciences, University of Chinese Academy of Sciences,
Beijing 100190, China}
\maketitle

\renewcommand{\thesection}{\Alph{section}} \renewcommand{\thefigure}{S%
\arabic{figure}} \renewcommand{\thetable}{S\Roman{table}} %
\setcounter{figure}{0} \renewcommand{\theequation}{S\arabic{equation}} %
\renewcommand{\thesection}{\Alph{section}} \renewcommand{\thefigure}{S%
\arabic{figure}} \renewcommand{\thetable}{S\Roman{table}}

\bigskip The notation used in this Supplemental Material follows that
introduced in the main text.

\section{Noise state for four photon case}

In this section, we predict a novel noise state by approximately solving the
following dynamical equation of the single-cavity retard Green's function:
\begin{equation}
i\dot{G}_{0}^{R}\left( t\right) =\delta \left( t\right) -\theta \left(
t\right) \left\langle \left[ \left( i\omega _{c}-\chi \right) a\left(
t\right) ,a^{\dag }\left( 0\right) \right] \right\rangle -\theta \left(
t\right) \left\langle \left[ \left( iU+\kappa \right) a^{\dag }a^{2}\left(
t\right) ,a^{\dag }\left( 0\right) \right] \right\rangle .  \label{DEQ}
\end{equation}%
By defining\
\begin{equation}
\langle \langle F\left( a\left( t\right) ,a^{\dag }\left( t\right) \right)
;a^{\dag }\left( 0\right) \rangle \rangle =-i\theta \left( t\right)
\left\langle \left[ F\left( a\left( t\right) ,a^{\dag }\left( t\right)
\right) ,a^{\dag }\left( 0\right) \right] \right\rangle ,  \label{F}
\end{equation}%
where $F\left( a\left( t\right) ,a^{\dag }\left( t\right) \right) $ is an
arbitrary function of $a\left( t\right) $ and $a^{\dag }\left( t\right) $,
we obtain a recursive relation about the high-order time-ordered correlation
functions in the real-time space as%
\begin{equation}
i\frac{d}{dt}\langle \langle a^{\dag l}a^{m}\left( t\right) ;a^{\dag }\left(
0\right) \rangle \rangle =\delta \left( t\right) \left\langle \left[ a^{\dag
l}a^{m}\left( t\right) ,a^{\dag }\left( 0\right) \right] \right\rangle
+\theta \left( t\right) \left\langle \left[ \frac{d}{dt}a^{\dag
l}a^{m}\left( t\right) ,a^{\dag }\left( 0\right) \right] \right\rangle ,
\label{Gmn}
\end{equation}%
where $l$\ and $m$\ are integers. According to the quantum regression
theorem \cite{Walls}, the second term on the right hand of Eq.~(\ref{Gmn})
depends on
\begin{equation}
\frac{d\left\langle a^{\dag l}a^{m}\left( t\right) \right\rangle }{dt}%
=A_{lm}\left\langle a^{\dag l}a^{m}\left( t\right) \right\rangle
+B_{lm}\left\langle a^{\dag l+1}a^{m+1}\left( t\right) \right\rangle
+C_{lm}\left\langle a^{\dag l-1}a^{m-1}\left( t\right) \right\rangle ,
\label{Dmn}
\end{equation}%
where
\begin{equation}
A_{lm}=i\omega _{c}\left( l-m\right) +i\frac{U}{2}[l\left( l-1\right)
-m\left( m-1\right) ]+\chi (l+m)-\frac{\kappa }{2}\left[ l\left( l-1\right)
+m\left( m-1\right) \right] ,
\end{equation}%
\begin{equation}
B_{lm}=iU\left( l-m\right) -\kappa \left( l+m\right) ,
\end{equation}%
\begin{equation}
C_{lm}=\gamma _{p}lm.
\end{equation}%
Equations (\ref{Gmn}) and (\ref{Dmn}) are our main results of the
incoherent-pumped dissipative single cavity. Solving them, we can capture
all information about the photon field distribution and the single-particle
excitation spectra. In general, we cannot obtain the full solutions of Eqs.~(%
\ref{Gmn}) and (\ref{Dmn}), since they depend crucially on the photon
number. In current experiments \cite{EIT}, the maximal photon number in the
cavity is small. In the following discussions, we assume that there are at most four
incident photons, and then obtain Eqs.~(\ref{Gmn}) and (\ref{Dmn}) after
neglecting the time-ordered correlation functions higher than the fourth
order.

We first qualitatively analyze the characteristics of Eq.~(\ref{Dmn}), which
determines the photon field distribution. By setting $d\left\langle a^{\dag
l}a^{m}\left( t\right) \right\rangle /dt=0$, we can capture the information
about all kinds of the one-time correlation functions of steady state. In
these correlation functions, the non-diagonal terms ($l\!\neq \!m$) reflect
the coupling between different photon Fock states \cite{Walls} and can be
used as the criterion of coherence between photons. To understand that, we
choose $l=0$\ and $m=1$, which leads to a steady equation of $\left\langle
a\right\rangle $, i.e., $\left( -i\omega _{c}+\chi \right) \left\langle
a\right\rangle =\left( iU+\kappa \right) \left\langle a^{\dag
}a^{2}\right\rangle $. This equation has a normal solution, with $%
\left\langle a\right\rangle =\left\langle a^{\dag }a^{2}\right\rangle =0$,
which describes an incoherent steady-state property of system. In addition,
the diagonal terms ($l=m$) determine the mean-photon number $\langle a^{\dag
}a\rangle $\ and its fluctuations \cite{Walls}. When $l=m=1$, we obtain a
steady equation of $\langle a^{\dag }a\rangle $\ as $2\chi \langle a^{\dag
}a\rangle =2\kappa \left\langle a^{\dag 2}a^{2}\right\rangle -\gamma _{p}$,
which contains a constant $\gamma _{p}$, originating from the different time
orders of $a$\ and $a^{\dag }$\ in the single-photon pumping and loss
processes. This constant reflects the environment-induced quantum
fluctuations of driven-dissipative process and makes $\langle a^{\dag
}a\rangle \neq 0$\ for any case. It should be emphasized that the above
properties still remain in the higher-order non-diagonal and diagonal terms.
Therefore, we find that the incoherent state, characterized by $\left\langle
a\right\rangle =0$\ and $\langle a^{\dag }a\rangle \neq 0$, is a \textit{%
noise} state, rather than a vacuum state, with $\left\langle a\right\rangle
=0$ and $\left\langle a^{\dag }a\right\rangle =0$, which can be derived from
the mean-field approximation (see the following section). The denomination
of \textquotedblleft noise" implies the dephasing of the photon field, with
random phase and amplitude.

For four incident photons considered here, we can truncate Eqs.~(\ref{Gmn})
and (\ref{Dmn}) to the fourth order. In such case, the complete expressions
of the truncated Eqs.~(\ref{Gmn}) and (\ref{Dmn}) in the frequency space are
given by
\begin{equation}
\left\{
\begin{array}{c}
\omega \langle \langle a;a^{\dag }\rangle \rangle _{\omega }=1+\left( \omega
_{c}+i\chi \right) \langle \langle a;a^{\dag }\rangle \rangle _{\omega
}+\left( U-i\kappa \right) \langle \langle a^{\dag }a^{2};a^{\dag }\rangle
\rangle _{\omega } \\
\omega \langle \langle a^{\dag }a^{2};a^{\dag }\rangle \rangle _{\omega
}=2\langle a^{\dag }a\rangle +\left( \omega _{c}+U+3i\chi -i\kappa \right)
\langle \langle a^{\dag }a^{2};a^{\dag }\rangle \rangle _{\omega } \\
+2i\gamma _{p}\langle \langle a;a^{\dag }\rangle \rangle _{\omega }+\left(
U-3i\kappa \right) \langle \langle a^{\dag 2}a^{3};a^{\dag }\rangle \rangle
_{\omega } \\
\omega \langle \langle a^{\dag 2}a^{3};a^{\dag }\rangle \rangle _{\omega
}=3\langle a^{\dag 2}a^{2}\rangle +\left( \omega _{c}+2U+5i\chi -4i\kappa
\right) \langle \langle a^{\dag 2}a^{3};a^{\dag }\rangle \rangle _{\omega }
\\
+6i\gamma _{p}\langle \langle a^{\dag }a^{2};a^{\dag }\rangle \rangle
_{\omega }+\left( U-5i\kappa \right) \langle \langle a^{\dag 3}a^{4};a^{\dag
}\rangle \rangle _{\omega } \\
\omega \langle \langle a^{\dag 3}a^{4};a^{\dag }\rangle \rangle _{\omega
}=4\langle a^{\dag 3}a^{3}\rangle +\left( \omega _{c}+3U+7i\chi -9i\kappa
\right) \langle \langle a^{\dag 3}a^{4};a^{\dag }\rangle \rangle _{\omega }
\\
+12i\gamma _{p}\langle \langle a^{\dag 2}a^{3};a^{\dag }\rangle \rangle
_{\omega }%
\end{array}%
\right. .  \label{TGmn}
\end{equation}%
\begin{equation}
\left\{
\begin{array}{c}
\frac{d\langle a\rangle }{dt}=\left( -i\omega _{c}+\chi \right) \langle
a\rangle -\left( iU+\kappa \right) \langle a^{\dag }a^{2}\rangle \\
\frac{d\langle a^{\dag }\rangle }{dt}=\left( i\omega _{c}+\chi \right)
\langle a^{\dag }\rangle -\left( -iU+\kappa \right) \langle a^{\dag
2}a\rangle \\
\frac{d\langle a^{\dag }a\rangle }{dt}=2\chi \langle a^{\dag }a\rangle
-2\kappa \langle a^{\dag 2}a^{2}\rangle +\gamma _{p} \\
\frac{d\langle a^{\dag }a^{2}\rangle }{dt}=\left( -i\omega _{c}-iU+3\chi
-\kappa \right) \langle a^{\dag }a^{2}\rangle -\left( iU+3\kappa \right)
\langle a^{\dag 2}a^{3}\rangle +2\gamma _{p}\langle a\rangle \\
\frac{d\langle a^{\dag 2}a\rangle }{dt}=\left( i\omega _{c}+iU+3\chi -\kappa
\right) \langle a^{\dag 2}a\rangle -\left( -iU+3\kappa \right) \langle
a^{\dag 3}a^{2}\rangle +2\gamma _{p}\langle a^{\dag }\rangle \\
\frac{d\langle a^{\dag 2}a^{2}\rangle }{dt}=\left( 4\chi -2\kappa \right)
\langle a^{\dag 2}a^{2}\rangle -4\kappa \langle a^{\dag 3}a^{3}\rangle
+4\gamma _{p}\langle a^{\dag }a\rangle \\
\frac{d\langle a^{\dag 2}a^{3}\rangle }{dt}=\left( -i\omega _{c}-2iU+5\chi
-4\kappa \right) \langle a^{\dag 2}a^{3}\rangle -\left( iU+5\kappa \right)
\langle a^{\dag 3}a^{4}\rangle +6\gamma _{p}\langle a^{\dag }a^{2}\rangle \\
\frac{d\langle a^{\dag 3}a^{2}\rangle }{dt}=\left( i\omega _{c}+2iU+5\chi
-4\kappa \right) \langle a^{\dag 3}a^{2}\rangle -\left( -iU+5\kappa \right)
\langle a^{\dag 4}a^{3}\rangle +6\gamma _{p}\langle a^{\dag 2}a\rangle \\
\frac{d\langle a^{\dag 3}a^{3}\rangle }{dt}=\left( 6\chi -6\kappa \right)
\langle a^{\dag 3}a^{3}\rangle -6\kappa \langle a^{\dag 4}a^{4}\rangle
+9\gamma _{p}\langle a^{\dag 2}a^{2}\rangle \\
\frac{d\langle a^{\dag 3}a^{4}\rangle }{dt}=\left( -i\omega _{c}-3iU+7\chi
-9\kappa \right) \langle a^{\dag 3}a^{4}\rangle +12\gamma _{p}\langle
a^{\dag 2}a^{3}\rangle \\
\frac{d\langle a^{\dag 4}a^{3}\rangle }{dt}=\left( i\omega _{c}+3iU+7\chi
-9\kappa \right) \langle a^{\dag 4}a^{3}\rangle +12\gamma _{p}\langle
a^{\dag 3}a^{2}\rangle \\
\frac{d\langle a^{\dag 4}a^{4}\rangle }{dt}=\left( 8\chi -12\kappa \right)
\langle a^{\dag 4}a^{4}\rangle +16\gamma _{p}\langle a^{\dag 3}a^{3}\rangle%
\end{array}%
\right. .  \label{TDmn}
\end{equation}%
Solving Eqs.~(\ref{TGmn}) and (\ref{TDmn}), we can obtain $G^{R}(\omega
)=\langle \langle a;a^{\dag }\rangle \rangle _{\omega }$. The characteristic
frequencies $\omega _{0}$ of the single-particle excitation spectra are the
poles of $G^{R}(\omega )$, i.e., $\left[ G^{R}(\omega _{0})\right] ^{-1}=0$.
In Fig.~(3) of the main text, we show the real and imaginary parts of the
characteristic frequencies $\omega _{0}$. We find that the single-particle
excitation spectra are divided into four branches. When the imaginary parts
of $\omega _{0}$ are negative, the noise state is stable, otherwise the
steady state is instable and a new coherent state emerges. In this coherent
state, all non-diagonal terms of Eq.~(\ref{Dmn}) are nonzero. And the
mean-field theory is more suitable to describe this case; see the following
section. We emphasize that in the noise state, the photons are localized in
each cavity, which is crucial for many-body steady-state phase transition,
as shown in the main text.

\section{Mean-field steady-state phase transition of single cavity}

In this section, we present a mean-field steady-state phase transition of
the incoherent-pumped dissipative single cavity. To do this, we first turn
Eq.~(\ref{DEQ}) into the frequency space, i.e.,
\begin{equation}
\omega \langle \langle a;a^{\dag }\rangle \rangle _{\omega }=1+\left( \omega
_{c}+i\chi \right) \langle \langle a;a^{\dag }\rangle \rangle _{\omega
}+\left( U-i\kappa \right) \langle \langle a^{\dag }a^{2};a^{\dag }\rangle
\rangle _{\omega },  \label{R1}
\end{equation}%
where $\langle \langle a;a^{\dag }\rangle \rangle _{\omega }$\ is the retard
Green's function in the frequency space and its inverse determines the
single-particle excitation spectra \cite{AB10}. This equation is not a
closed equation of $\langle \langle a;a^{\dag }\rangle \rangle _{\omega }$,
since it is decided by the second-order time-ordered correlation function $%
\langle \langle a^{\dag }a^{2};a^{\dag }\rangle \rangle _{\omega }$.
Obviously, we cannot directly derive the required single-particle excitation
spectra of single cavity from Eq.~(\ref{R1}), unless we take some
approximation.

In the spirit of the mean-field approximation, the photon distribution in
the cavity can be described by a classical photon field, characterized by a
pair of complex variables. Thus, we split the operators $a$ and $\,a^{\dag }$
into their steady-state expectation values and quantum fluctuations, i.e.,
\begin{equation}
a=\alpha +\delta \alpha \text{, \ \ }a^{\dag }=\alpha ^{\ast }+\delta \alpha
^{\dag },  \label{MF1}
\end{equation}%
where $\alpha =\left\langle a\right\rangle $ and $\alpha ^{\ast
}=\left\langle a^{\dag }\right\rangle $ are the mean-field amplitudes, and $%
\delta \alpha $ and $\delta \alpha ^{\dag }$ are the fluctuation operators.
The complex variables $\alpha $ and $\alpha ^{\ast }$ fully specify the
coherent properties of steady state and the mean-photon number can be
defined as $n=\left\vert \alpha \right\vert ^{2}$. Meanwhile, the stability
of steady state is decided by the dynamics of $\delta \alpha $ and $\delta
\alpha ^{\dag }$ \cite{Walls}. If only considering the quadratic term of
fluctuations in $\langle \langle a^{\dag }a^{2};a^{\dag }\rangle \rangle
_{\omega }$, we simplify Eq.~(\ref{R1}) as a closed equation,
\begin{equation}
\omega \langle \langle \delta \alpha ;\delta \alpha ^{\dag }\rangle \rangle
_{\omega }=1+\left( \omega _{c}+i\chi \right) \langle \langle \delta \alpha
;\delta \alpha ^{\dag }\rangle \rangle _{\omega }+\left( U-i\kappa \right)
\left\vert \alpha \right\vert ^{2}\langle \langle \delta \alpha ;\delta
\alpha ^{\dag }\rangle \rangle _{\omega },  \label{MF2}
\end{equation}%
where $\left\vert \alpha \right\vert ^{2}$\ should be determined
self-consistently. We assume that at the initial time the system is in the
vacuum state, with $\left\vert \alpha \right\vert ^{2}=0$, and then the
pumping is switched on adiabatically. In this case, the non-zero mean-field
amplitudes appear when the pumping strength crosses a critical value. This
critical point can be calculated by the dispersion relation of the vacuum
state, which is encoded in the poles of the inverse of the retard Green's
function, i.e.,
\begin{equation}
\omega =\omega _{c}+i\chi .  \label{MF3}
\end{equation}%
For $\chi <0$, the single-particle excitation decays in time and the vacuum
state is stable, whereas for $\chi >0$, the vacuum state becomes instable
and the fully dispersion relation is given by
\begin{equation}
\omega =\omega _{c}+U\left\vert \alpha \right\vert ^{2}+i(\chi -\kappa
\left\vert \alpha \right\vert ^{2}),  \label{MF4}
\end{equation}%
which has a steady solution, with $\left\vert \alpha \right\vert ^{2}=\chi
/\kappa $. Based on the above analysis, we find a mean-field steady-state
phase transition of the incoherent-pumped dissipative single cavity, from a
incoherent vacuum state, with $\left\langle a\right\rangle =0$ and $%
\left\langle a^{\dag }a\right\rangle =0$, to a coherent classical state,
with $\left\langle a\right\rangle \neq 0$ and $\left\langle a^{\dag
}a\right\rangle \neq 0$, at the critical point $\chi _{c}=0$. This phase
transition is independent of the photon-photon interaction. This mean-field
approximation can well describe the physical phenomena in the deep
coherent-state region. However, it cannot capture the behavior in the
critical region. The main reason is that this method has neglected the
important environment-induced quantum-fluctuation effect, governed by the
higher-order time-ordered correlation functions in the incoherent state; see
the main text for details.

\section{Mean-field many-body steady-state phase transition}

In this section, we present a mean-field many-body steady-state phase
transition of the 2D incoherent-pumped dissipative Bose-Hubbard model.

When replaced $a_{i}$ by $\alpha _{i}=\langle a_{i}\rangle $ in the 2D
Bose-Hubbard Hamiltonian (2) of the main text and assumed that all sites are
identical, i.e., $\alpha _{i}=\alpha _{j}$, the initial many-body problem is
reduced to a single-site problem, which is governed by the following
equation:
\begin{equation}
\frac{d\alpha }{dt}=\left[ -i\left( \omega _{c}-J+U\left\vert \alpha
\right\vert ^{2}\right) +\chi -\kappa \left\vert \alpha \right\vert ^{2}%
\right] \alpha .  \label{MFM1}
\end{equation}%
We can obtain self-consistent solutions of Eq.~(\ref{MFM1}) by assuming $%
\alpha =\alpha _{0}\exp \left[ -i(\omega _{c}-J+\mu )t\right] $, where $\mu
=U\left\vert \alpha _{0}\right\vert ^{2}$\ is introduced as an effective
chemical potential for dynamical stability \cite{Keeling}. The mean-photon
number $\left\vert \alpha \right\vert ^{2}$\ is derived from the real part
on the right-hand of Eq.~(\ref{MFM1}). For example, for $\chi <0$, Eq.~(\ref%
{MFM1}) only have a zero solution, with $\left\langle a\right\rangle
=\left\langle a^{\dag }\right\rangle ^{\ast }=\alpha _{0}=0$. This means
that the photon field have no coherent part and the system is in an
incoherent vacuum state, with $\left\langle a^{\dag }a\right\rangle
=\left\vert \alpha \right\vert ^{2}=0$. For $\chi >0$, there exists a
stationary solution, with $\left\vert \alpha _{0}\right\vert ^{2}=\chi
/\kappa $. In this case, the system is in a coherent classical state, with $%
\left\langle a\right\rangle =\left\langle a^{\dag }\right\rangle ^{\ast
}=a_{0}\exp \left( -i\mu t\right) /\sqrt{2}$ and $\left\langle a^{\dag
}a\right\rangle =\left\vert a_{cl}\right\vert ^{2}/2=\chi /\kappa $. In
conclusion, based on Eq.~(\ref{MFM1}), a mean-field many-body steady-state
phase transition of the 2D incoherent-pumped dissipative Bose-Hubbard model,
from the incoherent vacuum state to the coherent classical state, is
predicted at the critical point $\chi _{c}=0$.

\section{Many-body steady-state phase transition using the continuous-field
approximation}

In this section, we present a many-body steady-state phase transition of the
2D incoherent-pumped dissipative Bose-Hubbard model, using the
continuous-field approximation in the Keldysh functional-integral formalism

We first write a Keldysh partition function, which is fully equivalent to
the many-body Lindblad master equation (1) of the main text, as \cite{AB10}%
\begin{equation}
Z=\int \mathcal{D}[a_{+},a_{-}]\exp \left[ iS_{\text{K}}\left(
a_{+},a_{-}\right) \right] ,  \label{Z}
\end{equation}%
where $+$ and $-$ denote the forward and backward branches, and $S_{\text{K}%
} $ is the Keldysh action. By introducing the coherent (H) and dissipative
(D) terms, the Keldysh action is given by
\begin{equation}
S_{\text{K}}=S_{\text{H}}+S_{\text{D}},  \label{SK1}
\end{equation}%
where%
\begin{eqnarray}
S_{\text{H}} &=&\sum\limits_{i}\int_{t}\left\{ \left[ a_{i+}^{\dag }\left(
i\partial _{t}-\omega _{c}\right) a_{i+}-\frac{U}{2}a_{i+}^{\dag
}a_{i+}^{\dag }a_{i+}a_{i+}\right] -\left[ a_{i-}^{\dag }\left( i\partial
_{t}-\omega _{c}\right) a_{i-}-\frac{U}{2}a_{i-}^{\dag }a_{i-}^{\dag
}a_{i-}a_{i-}\right] \right\}  \notag \\
&&+\frac{J}{z}\sum\limits_{\left\langle i,j\right\rangle }\int_{t}\left[
a_{i+}^{\dag }a_{j+}-a_{i-}^{\dag }a_{j-}\right] ,  \label{SH}
\end{eqnarray}%
\begin{eqnarray}
S_{\text{D}} &=&-i\gamma _{p}\sum\limits_{i}\int_{t}\left[ a_{i+}^{\dag
}a_{i-}-\frac{1}{2}\left( a_{i+}a_{i+}^{\dag }+a_{i-}a_{i-}^{\dag }\right) %
\right] -i\gamma _{l}\sum\limits_{i}\int_{t}\left[ a_{i+}a_{i-}^{\dag }-%
\frac{1}{2}\left( a_{i+}^{\dag }a_{i+}+a_{i-}^{\dag }a_{i-}\right) \right]
\label{SD} \\
&&-i\kappa \sum_{i}\int_{t}\left\{ \left( a_{i+}a_{i-}^{\dag }\right) ^{2}-%
\frac{1}{2}\left[ \left( a_{i+}^{\dag }a_{i+}\right) ^{2}+\left(
a_{i-}^{\dag }a_{i-}\right) ^{2}\right] \right\} ,  \notag
\end{eqnarray}%
with $\int_{t}=\int dt$.

For a field-theory treatment of lattice models, we can use the continuous
approximation to study the long-distance behavior \cite{Maghrebi}. If $%
\left\langle a\right\rangle $ and $\left\langle a^{\dag }\right\rangle $\
are assumed to be small near the critical point, we expand the lattice
operators to second-order derivative in the spatial coordinate, i.e.,
\begin{equation}
a_{i}\mapsto a_{x},\text{ }a_{i+1}\mapsto a_{x}+\nabla a_{x}+\frac{1}{2}%
\nabla ^{2}a_{x}+\cdots ,  \label{FLC}
\end{equation}%
where we have made the lattice constant as $1$. As a result, the hopping
terms become
\begin{equation}
-\frac{J}{z}\sum_{\left\langle i,j\right\rangle }\left( a_{i}^{\dag
}a_{j}+a_{j}^{\dag }a_{i}\right) \mapsto -\int_{\mathbf{r}}\left( Ja_{%
\mathbf{r}}^{\ast }a_{\mathbf{r}}+\frac{J}{z}a_{\mathbf{r}}^{\ast }\nabla
^{2}a_{\mathbf{r}}\right) ,  \label{FTT}
\end{equation}%
with $\int_{\mathbf{r}}=\int dxdy$, and the coherent and dissipative Keldysh
actions turn into%
\begin{eqnarray}
S_{\text{H}} &=&\int_{t,\mathbf{r}}\left\{ \left[ a_{\mathbf{r}+}^{\ast
}\left( i\partial _{t}-\omega _{c}+J+\frac{J}{z}\nabla ^{2}\right) a_{%
\mathbf{r}+}-\frac{U}{2}a_{\mathbf{r}+}^{\ast }a_{\mathbf{r}+}^{\ast }a_{%
\mathbf{r}+}a_{\mathbf{r}+}\right] -\left[ a_{\mathbf{r}-}^{\ast }\left(
i\partial _{t}-\omega _{c}+J+\frac{J}{z}\nabla ^{2}\right) a_{\mathbf{r}%
-}\right. \right.  \notag \\
&&\left. \left. -\frac{U}{2}a_{\mathbf{r}-}^{\ast }a_{\mathbf{r}-}^{\ast }a_{%
\mathbf{r}-}a_{\mathbf{r}-}\right] \right\} ,  \label{SHX}
\end{eqnarray}%
\begin{eqnarray}
S_{\text{D}} &=&-i\gamma _{p}\int_{t,\mathbf{r}}\left[ a_{\mathbf{r}+}^{\ast
}a_{\mathbf{r}-}-\frac{1}{2}\left( a_{\mathbf{r}+}a_{\mathbf{r}+}^{\ast }+a_{%
\mathbf{r}-}a_{\mathbf{r}-}^{\ast }\right) \right] -i\gamma _{l}\int_{t,%
\mathbf{r}}\left[ a_{\mathbf{r}+}a_{\mathbf{r}-}^{\ast }-\frac{1}{2}\left(
a_{\mathbf{r}+}^{\ast }a_{\mathbf{r}+}+a_{\mathbf{r}-}^{\ast }a_{\mathbf{r}%
-}\right) \right]  \notag \\
&&-i\kappa \int_{t,\mathbf{r}}\left\{ \left( a_{\mathbf{r}+}a_{\mathbf{r}%
-}^{\ast }\right) ^{2}-\frac{1}{2}\left[ \left( a_{\mathbf{r}+}^{\ast }a_{%
\mathbf{r}+}\right) ^{2}+\left( a_{\mathbf{r}-}^{\ast }a_{\mathbf{r}%
-}\right) ^{2}\right] \right\} .  \label{SDX}
\end{eqnarray}

It is more convenient to discuss Eqs.~(\ref{SHX}) and (\ref{SDX}) in the
Keldysh basis,%
\begin{equation}
a_{cl}=\frac{a_{+}+a_{-}}{\sqrt{2}},\text{ }a_{q}=\frac{a_{+}-a_{-}}{\sqrt{2}%
},  \label{acaq}
\end{equation}%
where $a_{cl}$ and $a_{q}$ are the classical and quantum fields \cite{AB10},
and we ignore the index $\mathbf{r}$\ for simplicity. After a
straightforward calculation, the Keldysh action in the spatial coordinate is
rewritten as
\begin{equation}
S_{\text{K}}=\int_{t,\mathbf{r}}\left(
\begin{array}{cc}
a_{cl}^{\ast } & a_{q}^{\ast }%
\end{array}%
\right) \left(
\begin{array}{cc}
0 & P^{A} \\
P^{R} & i\left( \gamma _{p}+\gamma _{l}\right)%
\end{array}%
\right) \left(
\begin{array}{c}
a_{cl} \\
a_{q}%
\end{array}%
\right) -\frac{1}{2}\left[ \left( U+i\kappa \right) \left( a_{cl}^{\ast
2}a_{cl}a_{q}+a_{q}^{\ast 2}a_{cl}a_{q}\right) +\text{c.c.}\right] +2i\kappa
a_{cl}^{\ast }a_{cl}a_{q}^{\ast }a_{q},  \label{SK2}
\end{equation}%
where c.c.~denotes the complex conjugate, and the inverse retarded and
advanced single-particle Green's functions are given by
\begin{equation}
P^{R}=P^{A\dag }=i\partial _{t}-\omega _{c}+\frac{J}{2}\nabla ^{2}+J-i\chi .
\label{GF}
\end{equation}%
Eq.~(\ref{SK2}) is similar to that of Refs.~\cite{Sieberer2013,Sieberer2014}.

For the small quantum fluctuations, we employ the saddle-point
approximation,
\begin{equation}
\frac{\delta S_{\text{K}}}{\delta a_{cl}^{\ast }}=0  \label{Saddle1}
\end{equation}%
and$\;\;$%
\begin{equation}
\;\frac{\delta S_{\text{K}}}{\delta a_{q}^{\ast }}=0,  \label{Saddle2}
\end{equation}%
to find the stable points. Firstly, Eqs.~(\ref{Saddle1}) and (\ref{Saddle2})
lead to two equations
\begin{equation}
P^{A}a_{q}-\frac{1}{2}\left( U+i\kappa \right) \left\vert a_{c}\right\vert
^{2}a_{q}-\frac{1}{2}\left( U-i\kappa \right) \left( \left\vert
a_{q}\right\vert ^{2}a_{q}+a_{c}^{2}a_{q}^{\ast }\right) +2i\kappa
\left\vert a_{q}\right\vert ^{2}a_{c}=0,  \label{S1-s}
\end{equation}%
\begin{equation}
P^{R}a_{c}+i\left( \gamma _{p}+\gamma _{l}\right) a_{q}-\frac{1}{2}\left(
U+i\kappa \right) \left\vert a_{q}\right\vert ^{2}a_{c}-\frac{1}{2}\left(
U-i\kappa \right) \left( \left\vert a_{c}\right\vert
^{2}a_{c}+a_{q}^{2}a_{c}^{\ast }\right) +2i\kappa \left\vert
a_{c}\right\vert ^{2}a_{q}=0.  \label{S2-s}
\end{equation}%
In terms of Eq.~(\ref{S1-s}), we obtain a solution \cite{AB10}
\begin{equation}
a_{q}=0.  \label{AQ}
\end{equation}%
Substituting Eq.~(\ref{AQ}) into Eq.~(\ref{S2-s}) yields a steady-state
equation
\begin{equation}
i\frac{\partial a_{cl}}{\partial t}=\left( \omega _{c}-J-\frac{J}{2}\nabla
^{2}+\frac{U}{2}\left\vert a_{cl}\right\vert ^{2}\right) a_{cl}+i\left( \chi
-\frac{\kappa }{2}\left\vert a_{cl}\right\vert ^{2}\right) a_{cl}.
\label{MF}
\end{equation}

Based on the imaginary part on the right-hand of Eq.~(\ref{MF}), we can
obtain self-consistent solutions. For example, for $\chi <0$, Eq.~(\ref{MF})
only has a zero solution, with $a_{cl}=0$, and the mean value of photon
field is thus given by $\left\langle a\right\rangle =\left\langle a^{\dag
}\right\rangle ^{\ast }=a_{cl}/\sqrt{2}=0$. This implies that the photon
field has no coherent part, and the system is in an incoherent vacuum state,
with $\left\langle a^{\dag }a\right\rangle =\left\vert a_{cl}\right\vert
^{2}/2=0$. For $\chi >0$, there exists a stationary solution, with $%
a_{cl}=a_{0}\exp \left( -i\mu t\right) $, where $a_{0}=\left\vert
a_{cl}\right\vert $ and the parameter $\mu =U\left\vert a_{cl}\right\vert
^{2}/2$\ is also introduced as an effective chemical potential for dynamical
stability~\cite{Keeling}. In this case, the system is in a coherent
classical state, with $\left\langle a\right\rangle =\left\langle a^{\dag
}\right\rangle ^{\ast }=a_{0}\exp \left( -i\mu t\right) /\sqrt{2}$ and $%
\left\langle a^{\dag }a\right\rangle =\left\vert a_{cl}\right\vert
^{2}/2=\chi /\kappa $. The above results indicate that the positive
effective gain of single photon can pump a coherent photon field and is
balanced by the two-photon loss. Moreover, they demonstrate that a many-body
steady-state phase transition of the 2D incoherent-pumped dissipative
Bose-Hubbard model, from the incoherent vacuum state to the coherent
classical state, occurs at the critical point $\chi _{c}=0$.

However, the above results, similar to the mean-field predictions in
Sec.~(C), seem confusing, because they violate our intuitive experiences
about the strong correlation problems, where we expect the photon-photon
interaction in the same cavity and the photon hopping between adjacent
cavities play dominate roles in the critical behavior. A possible reason is
that we have assumed the environment-induced quantum fluctuations are small
to ensure the validity of the saddle-point approximation in Eqs.~(\ref%
{Saddle1}) and (\ref{Saddle2}). This approximation can be met only if the
system enters the deep coherent-state region, where the pumping is strong
and the coherent part of photons is large \cite{AB10}. It is not surprising
that the above method cannot make correct prediction about the critical
behavior. To overcome this shortcoming, the environment-induced quantum
fluctuations must be taken into account and a more effective method should
be established, which is the main purpose of this Letter. The detailed
discussions are shown in the main text as well as the following section.

\section{The non-equilibrium Dyson equation using non-equilibrium
strong-coupling expansion approximation}

To correctly capture the strong-correlated steady-state phase transition of
the 2D incoherent-pumped dissipative Bose-Hubbard model, here we develop a
non-equilibrium strong-coupling approximation in the Keldysh formalism. In
this section, we mainly obtain a non-equilibrium Dyson equation.

We assume that at the initial time every cavity is in the noise state, which
means that the photons are localized at each site. When increasing the
hopping strength, the photons can transition between adjacent cavities.
Therefore, we can present a strong-coupling expansion \cite{Metzner}, in
which all the single-site terms are regarded as the unperturbed parts and
the hopping terms are treated as perturbations.

In the Keldysh basis, the many-body partition function is written as%
\begin{equation}
Z=\int \mathcal{D}\left[ a_{cl},a_{q}\right] \exp [iS_{\text{0}}\left( a_{cl}%
\text{,}a_{q}\right) +iS_{\text{J}}\left( a_{cl}\text{,}a_{q}\right) ],
\label{Z0}
\end{equation}%
where $a_{cl}$ and $a_{q}$ are the classical and quantum fields as mentioned
above. The unperturbed Keldysh action in Eq.~(\ref{Z0}) is given by
\begin{equation}
S_{\text{0}}=\sum_{i}\int_{t}\left( a_{i,cl}^{\ast }\text{ }a_{i,q}^{\ast
}\right) \left(
\begin{array}{cc}
0 & \left[ G_{i,i}^{-1}\left( t\right) \right] ^{A} \\
\left[ G_{i,i}^{-1}\left( t\right) \right] ^{R} & \left[ G_{i,i}^{-1}\left(
t\right) \right] ^{K}%
\end{array}%
\right) \binom{a_{i,cl}}{a_{i,q}},  \label{S0}
\end{equation}%
where $\left[ G_{i,i}^{-1}\left( t\right) \right] ^{R}$, $\left[
G_{i,i}^{-1}\left( t\right) \right] ^{A}$, and $\left[ G_{i,i}^{-1}\left(
t\right) \right] ^{K}$ are the inverses of the on-site retard Green's
function [equal to $G_{0}^{R}\left( t\right) $], advanced Green's function $%
G_{i,i}^{A}\left( t\right) =i\theta \left( -t\right) \langle \lbrack
a_{i}\left( t\right) ,a_{i}^{\dag }\left( 0\right) ]\rangle $, and Keldysh
Green's function $G_{i,i}^{K}\left( t\right) =-i\langle \lbrack a_{i}\left(
t\right) ,a_{i}^{\dag }\left( 0\right) ]_{+}\rangle $, respectively. The
advanced Green's function $G_{i,i}^{A}\left( t\right) $ can be derived
straightforwardly from the relation $\left[ G_{i,i}^{A}\left( t\right) %
\right] =\left[ G_{i,i}^{R}\left( t\right) \right] ^{\dag }$, whereas the
Keldysh Green's function $G_{i,i}^{K}\left( t\right) $ can be, in principle,
obtained by using the similar steps as calculating $G_{i,i}^{R}\left(
t\right) $. Similar to $G_{i,i}^{R}\left( t\right) $, the explicit
expression of $G_{i,i}^{K}\left( t\right) $ is very complicate. However,
here we do not present $G_{i,i}^{A}\left( t\right) $ and $G_{i,i}^{K}\left(
t\right) $ explicitly, since they does not affect the main results under the
first-order approximation of the photon hopping that we apply below.

\begin{figure}[tp]
\includegraphics[width=9.cm]{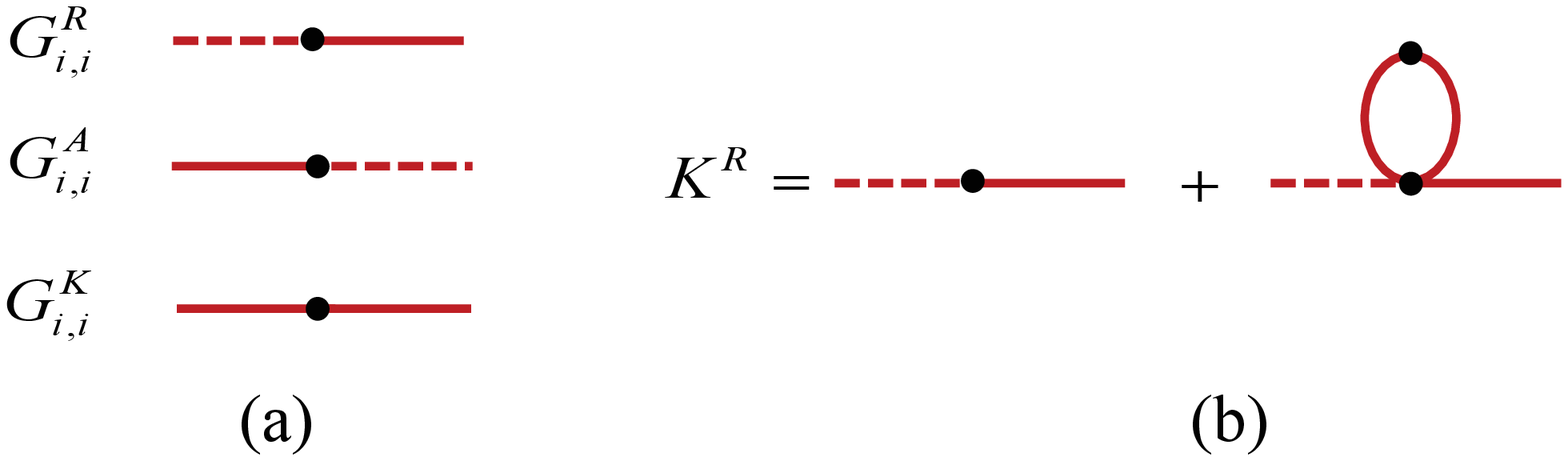}\newline
\caption{(Color online) (a) Graphic representation of the on-site retard
Green's function $G_{i,i}^{R}$, the advanced Green's function $G_{i,i}^{A}$,
and the Keldysh Green's function $G_{i,i}^{K}$. The solid and dashed lines
represent the classical and quantum fields $a_{cl}$ and $a_{q}$,
respectively. (b) The irreducible part of the retard Green's function $K^{R}$%
.}
\label{figs1}
\end{figure}

Since the above three unperturbed Green's functions arise from the many-body
Lndblad master equation (1) in the main text, they have a compact form
\begin{equation}
G_{i,i}^{\alpha \beta }\left( t\right) =\left(
\begin{array}{cc}
G_{i,i}^{K}\left( t\right) & G_{i,i}^{R}\left( t\right) \\
G_{i,i}^{A}\,\left( t\right) & 0%
\end{array}%
\right) =-i\left\langle a_{i,\alpha }(t)a_{i,\beta }^{\ast }(0)\right\rangle
_{S_{\text{0}}},  \label{RAK}
\end{equation}%
where $\alpha $, $\beta =\left( cl\text{, }q\right) $ and the average $%
\left\langle \cdots \right\rangle _{S_{\text{0}}}$ is taken with respect to
the unperturbed Keldysh action $S_{\text{0}}$ \cite{AB10}. Based on Eq.~(\ref%
{RAK}), we present a graphic representation of the three unperturbed Green's
functions; see Fig.~\ref{figs1}(a). The perturbation part of the Keldysh
action (\ref{Z0}) is given by
\begin{equation}
S_{\text{J}}=\sum_{\left\langle i,j\right\rangle }\int_{t}\frac{J_{ij}}{z}%
\left[ a_{i,cl}^{\ast }a_{j,q}+a_{i,q}^{\ast }a_{j,cl}+\text{c.c.}\right] .
\label{SJ}
\end{equation}%

To explore the many-body steady-state phase diagram, we should calculate the
full retard Green's function
\begin{equation}
\tilde{G}^{R}\left( t\right) =-i\left\langle a_{i,cl}(t)a_{i,q}^{\ast
}(0)\right\rangle _{S_{\text{0}}+S_{\text{J}}},  \label{GGF}
\end{equation}%
which is dressed by the hopping terms in the perturbed Keldysh action $S_{%
\text{J}}$. Since the unperturbed Keldysh action $S_{\text{0}}$ contains all
on-site terms, it is not a normal quadratic form on the field operators.
Thus, the Wick's theorem and the powerful perturbation technique, based on
the standard Feynman diagrams, cannot be applied. Instead, here we use a
non-equilibrium linked-cluster expansion approach in the Keldysh basis,
which gives a description of equilibrium or non-equilibrium strong
correlation systems within the same formalism \cite{AB10}. It has already
been used to study the Bose-Hubbard model with finite temperature \cite%
{Pelster} or time-varying hopping \cite{Kennett}. Following this method, we
sum an infinite set of diagrams by calculating the irreducible part of the
retard Green's function $K^{R}\left( t\right) $, which is connected, in the
frequency space, to the full retard Green's function via the following
equation:%
\begin{equation}
\tilde{G}^{R}\left( \mathbf{k},\omega \right) =\frac{K^{R}\left( \mathbf{k}%
,\omega \right) }{1-J\left( \mathbf{k}\right) K^{R}\left( \mathbf{k},\omega
\right) },  \label{GR}
\end{equation}%
with the 2D lattice dispersion $J(\mathbf{k})=2J\cos (\mathbf{k}\cdot
\mathbf{r})$, where $\mathbf{k}$ is the wave vector and $\mathbf{r}$ denotes
the lattice vector, with $\left\vert \mathbf{r}\right\vert =1$. In Fig.~\ref%
{figs1}(b), we show a chain diagram and a most relevant one-loop diagram of
the retard Green's function $K^{R}\left( \mathbf{k},\omega \right) $. The
chain diagram denotes the first-order approximation of the photon hopping
and the one-loop diagram represents the leading spatial related correction.
Numerical simulations have verified that this first-order approximation can
well describe the many-body steady-state properties of the 2D
non-equilibrium lattice systems \cite{SF15}, since in such systems the
higher-order corrections contain higher-order time correlation functions,
which decay very fast due to the coupling with environment \cite%
{Pelster,Kennett}. Thus, below we will mainly consider the chain diagram of
the retard Green's function $K^{R}\left( \mathbf{k},\omega \right) $.

By setting $K^{R}\left( \omega \right) =G_{i,i}^{R}\left( \omega \right) $
in Eq.~(\ref{GR}), we immediately obtain a non-equilibrium Dyson equation
about the inverse of the full retard Green's function in the momentum space,
i.e.,
\begin{equation}
\left[ \tilde{G}^{R}\left( \mathbf{k},\omega \right) \right] ^{-1}=\left[
G_{i,i}^{R}\left( \omega \right) \right] ^{-1}-\Sigma ^{R}\left( \mathbf{k}%
,\omega \right) ,  \label{DE}
\end{equation}%
where $\Sigma ^{R}\left( \mathbf{k},\omega \right) =J(\mathbf{k})$ is the
self-energy. Based on the non-equilibrium Dyson equation (\ref{DE}), we can
obtain the many-body steady-state phase diagram of the 2D incoherent-pumped
dissipative Bose-Hubbard model. The detailed discussions are shown in the
main text.